\documentclass[letterpaper, 10 pt, conference]{ieeeconf}  

\IEEEoverridecommandlockouts                              

\usepackage{amsmath,amssymb}
\usepackage{algorithm,algpseudocode}
\usepackage{graphicx}

\usepackage{cite}

\allowdisplaybreaks

\usepackage{subcaption} 

\usepackage{caption}
\usepackage{arydshln}

\title{\LARGE \bf
Hierarchical 2-degree-of-freedom control combining Youla–Kucera parameterization and model predictive control
}

\author{Zhiheng Zhao$^{1}$, Hans Henrik Niemann$^{1}$, John Bagterp Jørgensen$^{2}$
\thanks{$^{1}$Z. Zhao and H.H. Niemann are with Department of Electrical and Photonics Engineering, Technical University of Denmark, 2800 Kongens Lyngby, Denmark
        {\tt\small s232159@dtu.dk}, {\tt\small hahe@dtu.dk}}
\thanks{$^{2}$J.B. Jørgensen is with Department of Applied Mathematics and Computer Science, Technical University of Denmark, 2800 Kongens Lyngby, Denmark
        {\tt\small jbjo@dtu.dk}}%
}

\begin{document}

\maketitle
\thispagestyle{empty}
\pagestyle{empty}

\begin{abstract}
A hierarchical 2DOF (2-degree-of-freedom) structure combining Youla-Kučera (YK) parameterization and model predictive control (MPC) is presented in this paper. The YK parameterization employs the coprime factorization of the nominal system and controller, thereby introducing an auxiliary feedforward channel dedicated to system optimization and a controller parameterization channel. The feedforward channel is utilized to implement cascaded MPC for system optimization. The controller parameterization channel is utilized to achieve offset-free MPC by designing an appropriate YK parameter through the $\mathcal{H}_2$ optimal controller design.

\end{abstract}

\section{Introduction}
\label{sec:Introduction}

Developed in the 1970s, Youla-Kučera (YK) parametrization is a technique proposed by Youla et al. \cite{youla_modern_1976m, youla_modern_1976-1m}, and Kučera et al. \cite{KUCERA1975573}. By parameterizing the nominal controller with a stable YK parameter $Q$, a stable closed-loop system that satisfies specific performance criteria can be achieved. Consequently, system performance can be enhanced by simply modifying the YK parameter $Q$ to synthesize a parameterized controller $K(Q)$, circumventing the need for a complete controller redesign.
Conversely, the dual YK parameterization leverages dual YK parameter $S$ to parameterize the plant, characterizing the set of all linear time-invariant (LTI) plants stabilized by the nominal controller \cite{MAHTOUT202081}. The parameterized plant $G(S)$ can be employed to represent the model mismatch.

As a well-established control methodology, YK parameterization provides a systematic framework for diverse applications, as in Tay et al. \cite{tay_high_1998m}. These applications provide unique perspectives for achieving specific control objectives by utilizing YK parameter $Q$ and dual YK parameter $S$. While this structure provides an elegant perspective on nominal system parameterization, it remains inherently bound to classical control paradigms. Consequently, there is a strong motivation to integrate advanced control methodologies, e.g., model predictive control (MPC), which explicitly leverage future knowledge by predicting state evolution across a finite horizon and optimizing control inputs accordingly \cite{Jorgensen:2004,Jorgensen:etal:2004,Jorgensen:etal:2011,Maciejowski:2002}. However, the theoretical exposition of integrating such predictive algorithms within classical parameterized architectures remains relatively underdeveloped.

The combination of YK parameterization and MPC has been explored by researchers. Thomsen et al. \cite{thomsen_mpc_2008m} proposed an integrated YK–MPC approach that employs the dual YK parameter $S$ to explicitly capture model uncertainty, thereby enabling real-time corrections to the prediction equations within the MPC algorithm. Cheng et al. \cite{youla_chengqifeng} developed a robust MPC algorithm by embedding a predesigned YK parameter $Q$ to minimize the $\mathcal{H}_\infty$ norm of the nominal sensitivity function, retaining nominal performance while improving robustness. Fundamentally, these previous works adopt a unified approach where the YK parameterization is deeply embedded within the MPC predictive dynamics. Consequently, the optimization algorithm and the robustifying structure are heavily interdependent, lacking the flexibility of a decoupled design.

\begin{figure}[!t]
\centering
\includegraphics[width=0.8\columnwidth]{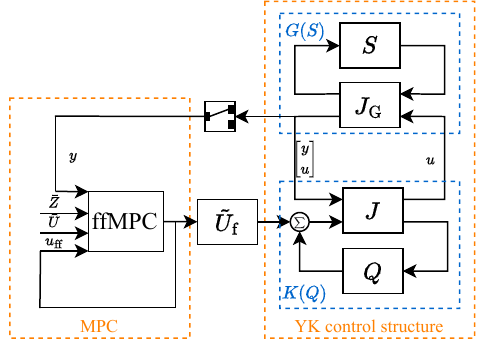}
\caption{2DOF control structure combining YK parameterization and MPC.}
\label{fig:overview}
\end{figure}

To address these limitations, this paper proposes a hierarchical architecture that structurally decouples YK parameterization and MPC, fully preserving the distinct advantages of both frameworks, as illustrated in Fig. \ref{fig:overview}. Within the inner YK control structure, the parameterized plant $G(S)$ is internally stabilized by the parameterized controller $K(Q)$. Specifically, the LTI systems $J_\text{G}$ and $J$ encapsulate the nominal mathematical models of the plant and controller, which are then parameterized by the dual YK parameter $S$ and the YK parameter $Q$, respectively. Meanwhile, the external MPC module acts as a feedforward optimizer, supplying the optimal control signal $u_{\text{ff}}$ based on the future output reference $\bar{Z}$ and control reference $\bar{U}$. Finally, $\tilde{U}_\text{f}$ serves as a bridging subsystem that maps the feedforward control signal into the YK control architecture.




In the YK control structure, the nominal plant is stabilized by a state-feedback gain derived from the decomposition of unconstrained MPC. This design choice is to preserve the primary feedback dynamics of the unconstrained MPC. The inner structure exhibits high computational efficiency since all constituent systems are LTI. Furthermore, by incorporating YK parameter $Q$, offset-free tracking can be realized. Combined with the feedforward MPC module (ffMPC), the hierarchical control structure simultaneously achieves optimal control performance and offset-free control.


The remainder of this paper is organized as follows. Section \ref{sec:MathematicalDerivations} presents the mathematical derivations leading to the formulation of the 2DOF YK parameterization control structure. Section \ref{sec:DesignFeedbackFeedforward} addresses the design of the optimization components, specifically the YK parameter $Q$ and the hierarchical MPC module. Finally, Section \ref{sec:CASE STUDY} evaluates the proposed control architecture through a case study.


\section{Mathematical derivations}
\label{sec:MathematicalDerivations}

\subsection{System description}
To demonstrate the mathematical derivations for the proposed hierarchical control strategy, the physical plant is formulated as a discrete LTI state-space model. The dynamics of the system are described as follows:
\begin{subequations}\label{eq:state space original pure expression}
    \begin{alignat}{3}
        x_{k+1} &&&= A x_k + B u_k + E d_k \\
        y_k &&&= C x_k
    \end{alignat}
\end{subequations}
where $x_k \in \mathbb{R}^{n_x}$, $u_k \in \mathbb{R}^{n_u}$, $y_k \in \mathbb{R}^{n_y}$, and $d_k \in \mathbb{R}^{n_d}$ represent the state, control input, measured output, and unmeasured exogenous disturbance at discrete time instant $k$, respectively. Matrices $A, B, C$, and $E$ are system matrices of appropriate dimensions. It is assumed that the system is stabilizable and detectable.

Throughout this paper, the unmeasured exogenous disturbance $d_k$ is assumed to be varying sufficiently slowly relative to the system dynamics.

\subsection{Decomposition of MPC algorithms}\label{Decomposition of MPC algorithms}
For a linear MPC, its optimization can be formulated as a linear-quadratic (LQ) problem. Taking the penalty terms of set-point tracking $\phi_z$, control input deviation $\phi_u$, and control input rate-of-change $\phi_{\Delta u}$, for example, the quadratic problem for an unconstrained MPC is given by:
\begin{subequations}
\begin{alignat}{5}
&\min_{x,u,z}\quad && \phi(x,u,z) = \phi_z + \phi_u  + &&\phi_{\Delta u},  \label{eq:mpc_obj}\\
&\text{s.t.}\,  && x_0 = \hat{x}_t, \label{eq:mpc_x0}\\
&                  && x_{k+1} = A x_k + B u_k, \quad && k = 0,1,\dots,N-1, \label{eq:mpc_dyn}\\
&                  && z_k = C x_k, \quad\;&&  k = 1,2,\dots,N. \label{eq:mpc_out}
\end{alignat}
\end{subequations}
Notably, the state-space model for prediction is given as follows, as disturbance is unknown:
\begin{equation}\label{eq:original system}
G : \left[\begin{array}{c|cccc}
A & B 
\\ \hline
C & 0 
\end{array}\right]
\end{equation}
The objective terms can be expanded as follows:
\begin{equation}
    \phi_j = \frac{1}{2}U_k^\top H_j U_k + g_j^\top U_k + \rho_j, \quad j \in \{z, u, \Delta u\}
\end{equation}
where:
\begin{subequations}
\begin{align}
    H_z &= (\bar W_z \Gamma)^\top(\bar W_z \Gamma)
    \\
     g_z & =   [(\bar W_z \Gamma)^\top\bar W_z(\Phi_x \hat x_t - \bar Z_k)] \label{eq: gz prediction}
     \\
     \rho_z &= \frac{1}{2}[\bar W_z(\Phi_x \hat{x}_t-\bar Z_k)]^\top
        [(\Phi_x \hat{x}_t-\bar Z_k)]
    \\
    H_u & = \bar W_u^\top \bar W_u
    \\
    g_u & = [-\bar W_u^\top \bar W_u \bar U_k]
    \\
    \rho_u & = \frac{1}{2} \bar U_k^\top \bar W_u^\top \bar W_u \bar U_k
    \\
    H_{\Delta u} & = (\bar W_{\Delta u}\Lambda)^\top(\bar W_{\Delta u}\Lambda)
    \\
    g_{\Delta u} &= 
            [-(\bar W_{\Delta u}\Lambda)^\top\bar W_{\Delta u}I_0 u_{k-1}]
    \\
    \rho_{\Delta u} &= \frac{1}{2}(\bar W_{\Delta u}I_0 u_{k-1})^\top(\bar W_{\Delta u}I_0 u_{k-1})
    \\
    \Phi_x &= \begin{bmatrix}
            (C A)^\top &
(C A^{2})^\top &
\cdots &
(C A^{N})^\top
        \end{bmatrix}^\top
    \\
    \Gamma &= \begin{bmatrix}
H_{1} & 0 & 0 & \cdots & 0\\
H_{2} & H_{1} & 0 & \cdots & 0\\
H_{3} & H_{2} & H_{1} & \cdots & 0\\
\vdots & \vdots & \vdots & \ddots & \vdots \\
H_{N} & H_{N-1} & H_{N-2} & \cdots & H_1
        \end{bmatrix}
    \\
    H_i &= CA^{i-1}B
    \\
    \Lambda &= \begin{bmatrix}
        I & 0 & 0 & \cdots &0 & 0\\
        -I & I &0 &\cdots &0 &0 \\
        0 & -I & I & \cdots & 0 & 0\\
        \vdots & \vdots & \vdots & \ddots & \vdots & \vdots \\
        0 & 0 & 0 & \cdots & -I & I
    \end{bmatrix} \label{eq: lambda matrix}
    \\
    I_0 & = \begin{bmatrix}
            I &
            0&
            0&
            \cdots &
            0
        \end{bmatrix}^\top \label{eq:extraction matrix math}
    \\
    \bar W_{\{z, u, \Delta u\}} &= I_{N\times N} \otimes W_{\{z, u, \Delta u\}} \label{eq:end}
\end{align} 
\end{subequations}
$W_z$, $W_u$ and $W_{\Delta u}$ are the weight matrices for the penalizing terms for $\phi_z$, $\phi_u$ and $\phi_{\Delta u}$, respectively. $\bar Z_k$ and $\bar U_k$ represent the stacked reference vectors for the output and control signal over the prediction horizon.

The solution for the unconstrained MPC in (\ref{eq:mpc_obj}) is when the derivative equals 0; that is:
\begin{subequations}
\begin{equation}
    \nabla \phi = HU_k + g = 0
\end{equation}
where
\begin{align}
    H &= H_z + H_u + H_{\Delta u} \\
    g &= g_z + g_u + g_{\Delta u}
\end{align}
\end{subequations}
The analytical solution for the optimal control sequence $U_k$ of the unconstrained quadratic problem is given by:
\begin{equation}
    \begin{split}
        U_k &= -H^{-1} g = - H^{-1} (g_z + g_u + g_{\Delta u}) \\
    &= -H^{-1}
        \Gamma^\top \bar W_z ^\top\bar W_z\Phi_x \hat x_{t}
        +
        {
        H^{-1}
        \Gamma^\top \bar W_z ^\top\bar W_z} \bar Z_k
        \\
        & \quad +
        {
        H^{-1}
        \bar W_u^\top \bar W_u} \bar U_k
        +
        {
        H^{-1}
        \Lambda^\top \bar W_{\Delta u}^\top \bar W_{\Delta u}I_0 } u_{k-1}
\end{split}
\end{equation}
In accordance with the receding horizon principle, only the first element of the optimal control sequence $U_k$ is applied to the plant. Consequently, the control law for the unconstrained MPC is given by:
\begin{equation}\label{eq:control law uncons}
    u_k = \underbrace {L_\text{x} \hat x_k}_{\text{feedback}} 
    + 
    L_{\bar Z} \bar Z_k + L_{\bar U}\bar U_k + L_{\Delta u} u_{k-1}
\end{equation}
where
\begin{subequations}
\begin{align}
    L_\text{x} &= -{I_{0,U_x}} H^{-1}
        \Gamma^\top \bar W_z ^\top\bar W_z\Phi_x 
        \\
    L_{\bar Z} &= {I_{0,U_x}}H^{-1}
        \Gamma^\top \bar W_z ^\top\bar W_z\\
    L_{\bar U} &= {I_{0,U_x}} H^{-1}
        \bar W_u^\top \bar W_u
    \\
    L_{\Delta u} &= I_{0,U_x}{
        H^{-1}
        \Lambda^\top \bar W_{\Delta u}^\top \bar W_{\Delta u}I_0 }
        \\
    I_{0,U_x}
    &=
    \begin{bmatrix}
        I & 0 &0 &\cdots & 0
    \end{bmatrix}
\end{align}
\end{subequations}
In (\ref{eq:control law uncons}), the state feedback term is extracted and employed as the nominal controller in YK parameterization. Although the term involving $u_{k-1}$ may also be interpreted as a feedback signal, established classical architecture lacks the structural flexibility to directly accommodate this dynamic.


Incorporating the feedback gain derived from the MPC decomposition allows the nominal controller within the YK structure to capture the main stabilizing feedback dynamics of the MPC.


\subsection{State-space model of nominal plant and controller}\label{State-space model of nominal plant and controller}
Since the feedback component requires the current state $x_k$, a stationary Kalman filter is employed to yield the estimate $\hat{x}_{k|k}$. This stationary formulation ensures the nominal controller admits the LTI state-space realization required for coprime factorization, with its state-space form given by:
\begin{equation}\label{eq: youla K block ss form brief}
K_n:\left[\begin{array}{c|cccc}
(I-K_{\text{fx}}C)A & K_{\text{fx}} & (I-K_{\text{fx}}C)B
\\ \hline
L_\text{x}(I-K_{\text{fx}}C)A & L_\text{x}K_{\text{fx}} & L_\text{x}(I-K_{\text{fx}}C)B
\end{array}\right]
\end{equation}
where $K_{\text{fx}}$ is the Kalman gain.



Notably, the input to the nominal controller $K_n$ is partitioned into distinct channels for the measurement $y_k$ and the control input $u_k$. This partition is necessary because the feedforward control signal $u_{\text{ff}}$ must be superimposed onto the feedback signal $u_{\text{fb}}$ to form the total control input $u$, ensuring accurate state estimation.



To ensure structural compatibility with the nominal controller $K_n$, the state-space form for the original plant $G$ in (\ref{eq:original system}) is augmented with an additional output channel for control input $u$, to formulate the nominal plant $G_{n}$ for coprime factorization, given by:
\begin{equation}\label{eq: youla G block ss form brief}
G_n:
\left[\begin{array}{c|cccc}
A & B 
\\ \hline
C & 0 
\\
0 & I 
\end{array}\right]
\end{equation}

With this structural alignment, $G_n$ and $K_n$ are well-posed for the subsequent coprime factorization.



\subsection{Derivations of YK parameterization}
The concept of YK parameterization can be illustrated as follows. Consider a stable plant $G_s\in \mathcal{RH}_\infty$ with stable feedback control $K_s\in \mathcal{RH}_\infty$, the closed-loop system $T_{\text{cls}}$ is given by \cite{MACIEJOWSKI202666}:
\begin{equation}
    T_{\text{cls}} = G_sK_s(I+G_sK_s)^{-1}
\end{equation}
Its corresponding YK parameter $Q_s$ is defined as:
\begin{equation}\label{eq: Q K expression}
    Q_s = K_s(I+G_sK_s)^{-1}
\end{equation}
As a result, the closed-loop transfer matrix reduces to $T_{\text{cls}} = G_sQ_s$. A key advantage of this parameterization is that closed-loop internal stability is ensured, provided that $Q_s$ remains stable, without the need to explicitly re-verify closed-loop stability with the nominal plant $G_s$.


If the assumption of stable controlled plant $G_s$ is relaxed, stable coprime factorization will be needed. Taking $G_n$ from (\ref{eq: youla G block ss form brief}) and $K_n$ from (\ref{eq: youla K block ss form brief}) for stable coprime factorization yields the following relations: 
\begin{subequations}
\begin{align}
        &G_n = NM^{-1} = \tilde M\tilde N^{-1}
    \\
    &K_n = UV^{-1} = \tilde V\tilde U^{-1}
\end{align}
\end{subequations}
In the above equations, subsystems $M$, $N$, $U$, $V$, $\tilde M$, $\tilde N$, $\tilde U$, and $\tilde V$ are all stable and proper transfer function matrices that are stabilized by cancellable denominators, and they satisfy the double Bezout equation \cite{tay_high_1998m}. One solution can be referred to as the state-space solution described in (\ref{eq:coprime fac}) in the appendix. By subpartitioning the state-space solution, subsystems can be extracted.

Using the subsystems above, the parameterized controller $K(Q)$ is given by \cite{tay_high_1998m}:
\begin{equation}\label{eq:introyoula parameterized KQ expamsion}
    \begin{split}
        K(Q) &= (U+MQ)(V+NQ)^{-1}
        \\
        & = K_n + \tilde V^{-1}Q(I+V^{-1}NQ)^{-1}V^{-1}
    \end{split}
\end{equation}
or using a lower linear fractional transformation (LFT) description \cite{ZhouDoyle_1998_EssentialsOfRobustControl}:
\begin{equation}
    K(Q) = \text{LFT}_l(J,Q)
\end{equation}
where $J$ is given by:
\begin{equation}
J = \begin{bmatrix}
        K_n & \tilde V^{-1} \\
        V^{-1} & -V^{-1}N
    \end{bmatrix}
\end{equation}



Extending the parameterization to a 2DOF structure requires incorporating the feedforward signal. By augmenting the nominal system and invoking (\ref{eq:introyoula parameterized KQ expamsion}), the overall controller can be decoupled into feedforward and feedback elements, yielding the augmented block $\mathcal{J}$ \cite{tay_high_1998m}:
\begin{equation}\label{eq:intrompc aug J block}
    \begin{split}
        \mathcal{J} &= \begin{bmatrix}
        K_{\text{ff}} & K_n & \tilde{V}^{-1}
        \\
        -V^{-1}N \tilde U_\text{f} & V^{-1} & -V^{-1}N
    \end{bmatrix}
    \\
    &=
    \begin{bmatrix}
        \begin{bmatrix}
        K_{\text{ff}}
        \\
        -V^{-1}N \tilde U_\text{f} 
    \end{bmatrix}
    & J
    \end{bmatrix}
    \end{split}
\end{equation}
where
\begin{equation}\label{eq:Uf}
    \tilde U_\text{f} = \tilde V K_{\text{ff}} 
\end{equation}
and $K_{\mathrm{ff}}$ denotes the LTI transfer matrix of the feedforward control module.

Similarly, the dual YK parameter can be employed for plant parameterization, which is given by:
\begin{equation}\label{eq:parametrization of GS}
    \begin{split}
        G(S) & = (N+VS)(M+US)^{-1}\\
        & = G_n + \tilde M^{-1}S(I+M^{-1}US)^{-1}M^{-1}
    \end{split}
\end{equation}
or using the upper LFT description:
\begin{equation}
    G(S) = \text{LFT}_u(J_\text{G},S)
\end{equation}
where $J_\text{G}$ is given by:
\begin{equation}\label{eq:introyoula paramterization structure JG}
    J_\text{G} = \begin{bmatrix}
        -M^{-1}U & M^{-1}\\
        \tilde M^{-1} & G_n
    \end{bmatrix}
\end{equation}

To incorporate the effects of exogenous signals (e.g., noise and disturbances), the plant model is augmented following the logic used for the nominal controller. The resulting augmented block $\mathcal{J}_{\mathrm{G}}$ is given by:
\begin{equation}
    \begin{split}
        \mathcal{J_\text{G}} &= \begin{bmatrix}
        -M^{-1}U & M^{-1} & -M^{-1}U \tilde N_d
        \\
        \tilde M^{-1} & G_n & G_d
    \end{bmatrix}
    \\
    & = 
    \begin{bmatrix}
    J_\text{G} & 
    \begin{bmatrix}
        -M^{-1}U \tilde N_d
        \\
        G_d
    \end{bmatrix}
    \end{bmatrix}
    \end{split}
\end{equation}
where
\begin{equation}
    \tilde{N}_d = \tilde{M}G_d
\end{equation}
and $G_d$ characterizes the exogenous signals such as disturbance and noise.

\begin{figure}[!t]
\centering
\includegraphics[width=1\columnwidth]{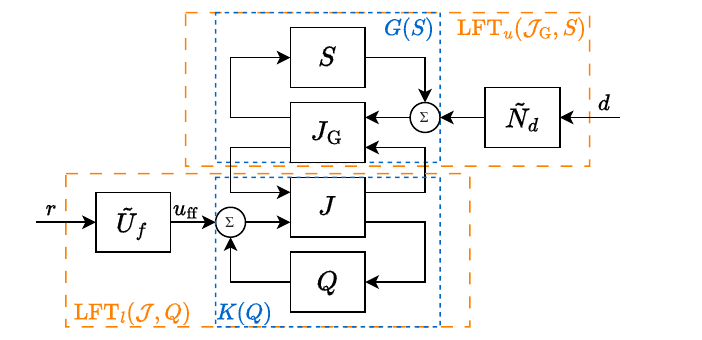}
\caption{Parameterized plant and controller control structure.}
\label{fig:parameterized plant and controller}
\end{figure}

In summary, the diagram of the parameterized YK control structure is shown in Fig. \ref{fig:parameterized plant and controller}. Its closed-loop stability is readily verified.
\section{Feedback and feedforward design}
\label{sec:DesignFeedbackFeedforward}


\subsection{Design of YK parameter}\label{Design of YK parameter}

\begin{figure}[!t]
\centering
\includegraphics[width=0.7\columnwidth]{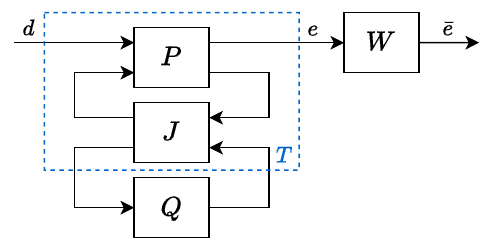}
\caption{Block diagram for designing YK parameter.}
\label{fig:Q design}
\end{figure}

Since the exact model mismatch is typically unavailable during the offline design phase, the dual YK parameter is initially set to zero ($S = 0$), causing the augmented block $J_G$ to reduce to the nominal plant. To facilitate the $\mathcal{H}_2$ control synthesis, an error output channel $e$ is introduced through matrix $C$. The corresponding generalized plant $P$ is thus formulated as:
\begin{equation}
P = \left[\begin{array}{c|c:c}
    A & E & B \\ \hline
    C & 0 & 0 \\ \hdashline
    C & 0 & 0 \\
    0 & 0 & I
\end{array}\right]
\end{equation}
The block diagram for the YK parameter design is depicted in Fig. \ref{fig:Q design}. The dashed box $T$ represents the closed-loop transfer matrix mapping the exogenous disturbance $d$ to the performance error $e$. $W$ denotes the performance weighting filter, which penalizes the error $e$ in specific frequency bands to achieve the desired disturbance rejection capabilities.

Notably, as the 2DOF architecture decouples reference tracking from disturbance rejection, the reference signal $r$ is intentionally omitted in Fig. \ref{fig:Q design}. Consequently, the system output is solely driven by the exogenous disturbance $d$. By minimizing the closed-loop $\mathcal{H}_2$ norm from $d$ to the filtered error $\bar e$, the effect of this disturbance can be attenuated.

The closed-loop transfer matrix $T$ can be obtained using lower LFT, i.e., $\text{LFT}_l(P,J)$:
\begin{equation}\label{eq:introyoula block t}
    T = \begin{bmatrix}
        P_{11} + P_{12}M\tilde U P_{21} & P_{12}M
        \\
        \tilde MP_{21}& 0
    \end{bmatrix} \overset{\text{def}}{=}
    \begin{bmatrix}
        T_{11} & T_{12}
        \\ T_{21}& 0
    \end{bmatrix}
\end{equation}
As disturbance is mainly interpreted as low-frequency signals,  a low-pass weighting filter matrix $W$ is introduced to actively penalize these low-frequency effects. By leveraging $\mathcal{H}_2$ control synthesis, the YK parameter $Q$ can be systematically determined. The design objective is formulated as finding a stable transfer matrix $Q \in \mathcal{RH}_\infty$ that minimizes the following cost function:
\begin{equation}
    \min_{Q\in \mathcal{RH_\infty}} ||W(T_{11} + T_{12}QT_{21})||_2^2
\end{equation}


Incorporating the designed parameter $Q$ into the structure in Fig. \ref{fig:parameterized plant and controller} yields the complete YK control framework.

\subsection{Reference tracking feedforward module design} \label{Reference tracking feedforward component design}
The feedforward module can be used to track the reference directly. To match the reference $r$ and output $y$, the static feedforward gain $K_{\mathrm{ff}}$ is computed by evaluating the inverse of the closed-loop DC gain, assuming $C(I-A-BL_{x})^{-1}B$ is invertible:
\begin{equation}\label{eq:ff ref}
    K_{\text{ff}} = [C(I-A-BL_{x})^{-1}B]^{-1}
\end{equation}
Subsequently, by calculating $\tilde{U}_{\mathrm{f}}$ via (\ref{eq:Uf}) and integrating it into the structure depicted in Fig. \ref{fig:parameterized plant and controller}, the 2DOF reference tracking configuration is established.

\subsection{MPC feedforward module design}\label{MPC as feedforward component}
In section \ref{Decomposition of MPC algorithms}, the unconstrained MPC formulation is decomposed into a linear state-feedback term, etc. However, in the presence of physical constraints, MPC no longer yields a closed-form solution but generates a numerical optimal control sequence at each sampling instant. By treating this dynamically computed MPC output as an exogenous reference signal, it can be seamlessly injected into the YK control structure.

As the disturbance and model mismatch are handled by YK parameter $Q$, the MPC module can focus on the dynamics of the nominal closed-loop system. By substituting the dynamic matrix $A$ with the closed-loop matrix $A_{\text{cl}}\overset{\text{def}}{=}A+BL_{\mathrm{x}}$ within the optimization problem from (\ref{eq:mpc_obj}) to (\ref{eq:end}), the feedforward MPC component is established. 

Constraints can also be implemented. Take constraints for states $x$, output $y$, control input $u$, and its change of rate $\Delta u$. The equivalent model of state evolution from the feedforward MPC's perspective is given by:
\begin{equation}
    x_{k+1} = \underbrace{(A+BL_{\mathrm{x}})}_{A_{\text{cl}}} x_k + \underbrace{BK_{\text{ff}}}_{B_{\text{cl}}}r_k
\end{equation}
In the predictive horizon $N$, the future state predictions can be formulated as:
\begin{equation}
    \bar{X}_k = \Phi_{\mathrm{cl}} x_k + \Gamma_{\mathrm{cl}} \bar{R}_k
\end{equation}
where 
\begin{subequations}
    \begin{alignat}{3}
        \bar{X}_k &= [x_{k+1|k}^\top, \dots, x_{k+N|k}^\top]^\top
        \\
        \bar{R}_k &= [r_{k|k}^\top, r_{k+1|k}^\top, \dots, r_{k+N-1|k}^\top]^\top
        \\
        \Phi_{\mathrm{cl}} &= \left[ (A_{\mathrm{cl}})^\top \ (A_{\mathrm{cl}}^2)^\top \ \dots \ (A_{\mathrm{cl}}^N)^\top \right]^\top
        \\
        \Gamma_{\mathrm{cl}} &= \begin{bmatrix}
    H_{\mathrm{cl},1} & 0 & \dots & 0 \\
    H_{\mathrm{cl},2} & H_{\mathrm{cl},1} & \dots & 0 \\
    \vdots & \vdots & \ddots & \vdots \\
    H_{\mathrm{cl},N} & H_{\mathrm{cl},N-1} & \dots & H_{\mathrm{cl},1}
\end{bmatrix}
\\
H_{\mathrm{cl},i} &= A_{\mathrm{cl}}^{i-1}B_{\mathrm{cl}}
    \end{alignat}
\end{subequations}
To enforce operational constraints on the state $x$ and output $y$, the following relationships are employed:
\begin{subequations}
\begin{gather}
    \bar{X}_{\min} \le \Phi_{\mathrm{cl}} x_k + \Gamma_{\mathrm{cl}} \bar{R}_k \le \bar{X}_{\max}
        \\
        \bar{Y}_{\min} \le \bar{C} (\Phi_{\mathrm{cl}} x_k + \Gamma_{\mathrm{cl}} \bar{R}_k) \le \bar{Y}_{\max}
\end{gather}
\end{subequations}
where $\bar{X}_{\{\min, \max\}}$, $\bar{Y}_{\{\min, \max\}}$ are the boundary requirements for the state $x$ and output $y$, and $\bar{C} = I_{N \times N} \otimes C$.

However, the actual control signal applied to the physical plant is $u_k = L_{\mathrm{x}}x_k + K_{\mathrm{ff}}r_k$. The predictive control sequence is given by:
\begin{equation}\label{eq:control constraints detive Uk}
    \bar{U}_k = \bar{L}_{\mathrm{x}} \bar{X}_{0,k} + \bar{K}_{\mathrm{ff}} \bar{R}_k = \bar{L}_{\mathrm{x}} \Phi_{\mathrm{cl},0} x_k + (\bar{L}_{\mathrm{x}} \Gamma_{\mathrm{cl},0} + \bar{K}_{\mathrm{ff}}) \bar{R}_k
\end{equation}
where 
\begin{subequations}
    \begin{gather}
        \bar{X}_{0,k} = [x_{k|k}^\top, \dots, x_{k+N-1|k}^\top]^\top = \Phi_{\mathrm{cl},0} x_k + \Gamma_{\mathrm{cl},0} \bar{R}_k
        \\
        \Phi_{\mathrm{cl},0} = \begin{bmatrix} I \\ \Phi_{\mathrm{cl}}^{(1:N-1)} \end{bmatrix}, \qquad
    \Gamma_{\mathrm{cl},0} = \begin{bmatrix} 0 \\ \Gamma_{\mathrm{cl}}^{(1:N-1)} \end{bmatrix}
    \end{gather}
\end{subequations}
In this equation, $\Phi_{\mathrm{cl}}^{(1:N-1)}$ and $\Gamma_{\mathrm{cl}}^{(1:N-1)}$ denote the submatrices comprising the first $N-1$ block rows of $\Phi_{\mathrm{cl}}$ and $\Gamma_{\mathrm{cl}}$, respectively. And $\bar{L}_{\mathrm{x}} = I_{N \times N} \otimes L_{\mathrm{x}}$, $\bar{K}_{\mathrm{ff}} = I_{N \times N} \otimes K_{\mathrm{ff}}$.

If constraints applied for the amplitude of the control signal, it becomes:
\begin{equation}
    \bar{U}_{\min} - \bar{L}_{\mathrm{x}} \Phi_{\mathrm{cl},0} x_k \le (\bar{L}_{\mathrm{x}} \Gamma_{\mathrm{cl},0} + \bar{K}_{\mathrm{ff}}) \bar{R}_k \le \bar{U}_{\max} - \bar{L}_{\mathrm{x}} \Phi_{\mathrm{cl},0} x_k
\end{equation}
where $\bar{U}_{\{\min, \max\}}$ is the boundary requirement for the control signal $u$.

By (\ref{eq:control constraints detive Uk}), the change of rate constraints for $\Delta u$ can also be derived, given by:
\begin{equation}
    \begin{split}
        \Delta \bar{U}_{\min} \le \Lambda \big[ \bar{L}_{\mathrm{x}} \Phi_{\mathrm{cl},0} x_k + (\bar{L}_{\mathrm{x}} \Gamma_{\mathrm{cl},0} + \bar{K}_{\mathrm{ff}}) \bar{R}_k \big] - I_0 u_{k-1} 
        \\
        \le \Delta \bar{U}_{\max}
    \end{split}
\end{equation}
where matrix $\Lambda$ and the extraction matrix $I_0$ are defined in (\ref{eq: lambda matrix}) and (\ref{eq:extraction matrix math}), respectively. From the discussions above, constraints can be enforced within the feedforward MPC module. 

Simply by replacing the reference $r$ as the first vector of $\bar{R}_k$ provided by feedforward MPC, MPC with constraints can be incorporated into the YK control structure. By enforcing constraints, the reference signal generated by the feedforward MPC can be bounded. Since the internal stability of the YK control structure can be guaranteed a priori, this bounded exogenous input inherently ensures that the system output remains bounded, thereby satisfying the Bounded-Input Bounded-Output (BIBO) stability.

Interestingly, if the feedforward MPC module fails to operate as expected, the feedforward signal can be switched to the control scheme described in \ref{Reference tracking feedforward component design}. While the optimality is compromised, the controlled system can still maintain accurate reference tracking.




\subsection{MPC with augmented Kalman filter}\label{sec:MPC with augmented Kalman filter}
To evaluate the disturbance rejection performance, traditional offset-free MPC formulation incorporating an integrator is briefly introduced.
By augmenting the model of the plant with the state of estimated disturbance $\hat d$, the augmented model is formulated as \cite{PANNOCCHIA2015342}:
\begin{subequations}
    \begin{align}
        \begin{bmatrix}
            \hat x_{k+1}\\\hat d_{k+1}
        \end{bmatrix}
        &=
        \underbrace{\begin{bmatrix}
            A & E \\ 0 & I
        \end{bmatrix}}_{A_a}
        \underbrace{\begin{bmatrix}
            \hat x_{k}\\\hat d_{k}
        \end{bmatrix}}_{x^a_k}
        +
        \underbrace{\begin{bmatrix}
            B \\ 0
        \end{bmatrix}}_{B_a}u_k
        \\
        y_k &= \underbrace{\begin{bmatrix}
            C & 0
        \end{bmatrix}}_{C_a} x^a_k
    \end{align}
\end{subequations}


To achieve offset-free tracking, the estimated disturbance $\hat{d}$ should be incorporated into the MPC prediction model. Under the assumption that the unmeasured disturbance remains constant throughout the prediction horizon, the linear objective term in (\ref{eq: gz prediction}) is reformulated as follows:
\begin{equation}
    g_z^\top  =   [(\bar W_z \Gamma)^\top\bar W_z(\Phi_x \hat x_t + \Phi_d \hat d_k- \bar Z_k)]^\top
\end{equation}
where
\begin{subequations}
    \begin{align}
        \Phi_d & = \begin{bmatrix}
        (CX_1E)^\top & (CX_2E)^\top & (CX_3E)^\top &\cdots
    \end{bmatrix}^\top
    \\
    X_n & = \sum_{i = 1}^n A^{i-1}
    \end{align}
\end{subequations}
The revised MPC algorithm can reject unknown disturbance with the augmented Kalman filter.
\section{CASE STUDY} \label{sec:CASE STUDY}

\subsection{Setup for simulation}
To test the proposed methods, a four-tank system is obtained as an example \cite{Christensen:etal:arXiv2025}.
Detailed derivations for the model and symbol meaning can be referred to in \cite{845876}. The parameters of the example are given in Table \ref{tab: parameters}, where $q_{1,2,3,4}$ are the flow rates, $a_{1,2,3,4}$ and $A_{1,2,3,4}$ represent the outlet and tank cross-sectional areas, respectively. The relationship between the variables is given by:
\begin{equation}
    q_{i} = \frac{a_{i}}{A_{i}}\sqrt{2gh_{i}},\quad i = 1,2,3,4
\end{equation}
where $g$ is the gravitational acceleration, and $h_{1,2,3,4}$ are liquid level in corresponding tank. $k_{1,2}$ are proportional constants relating the control voltage to the flow rate. $k_c$ is a calibration constant used for liquid-level measurement. $\gamma_{1,2}$ are the valve proportion constants. The superscript $0$ denotes the linearization point. The plant is discretized with sampling time $T = 1[s]$. For simplicity, the unknown disturbance gain matrix $E$ is equal to $B$.


\begin{table}[!t]
    \centering
    \caption{Parameter values of the four-tank system from \cite{845876}}
    \label{tab: parameters}
    \begin{tabular}{cc|cc|cc|cc}
        \hline
        Param. & Val. & Param. & Val. & Param. & Val. & Param. & Val.
        \\
        \hline
        $A_1$ & 28.0 & $A_2$ & 32.0 & $A_3$ & 28.0 & $A_4$ & 32.0
        \\
        $a_1$ & 0.071 & $a_2$ & 0.057 & $a_3$ & 0.071 & $a_4$ & 0.057
        \\
        $h_1^0$ & 12.4 & $h_2^0$ & 12.7 & $h_3^0$ & 1.8 & $h_4^0$ & 1.4
        \\
        $k_c$ & 0.5 & $g$ & 981 & $k_1$ & 3.33 & $k_2$ & 3.35
        \\
        $\gamma_1$ & 0.7 & $\gamma_2$ & 0.6 & $u_1^0$ & 3.00 & $u_2^0$ & 3.00
        \\
        \hline
    \end{tabular}
\end{table}

The low-pass filter $W$ is chosen to be:
\begin{equation}
    W = \begin{bmatrix}
        \frac{0.05}{z - 0.95}&0
        \\
        0&\frac{0.05}{z - 0.95}
    \end{bmatrix}
\end{equation}
Formulating the closed loop as in \ref{Design of YK parameter}, and using the $\mathcal{H}_2$ controller design, the YK parameter $Q$ can then be obtained.

Though the control signals are calculated based on the linear structure, the simulations are conducted based on the nonlinear four-tank system model.

\subsection{Simulation results and analysis}
\begin{figure*}[t]
\centering
\includegraphics[width=2\columnwidth]{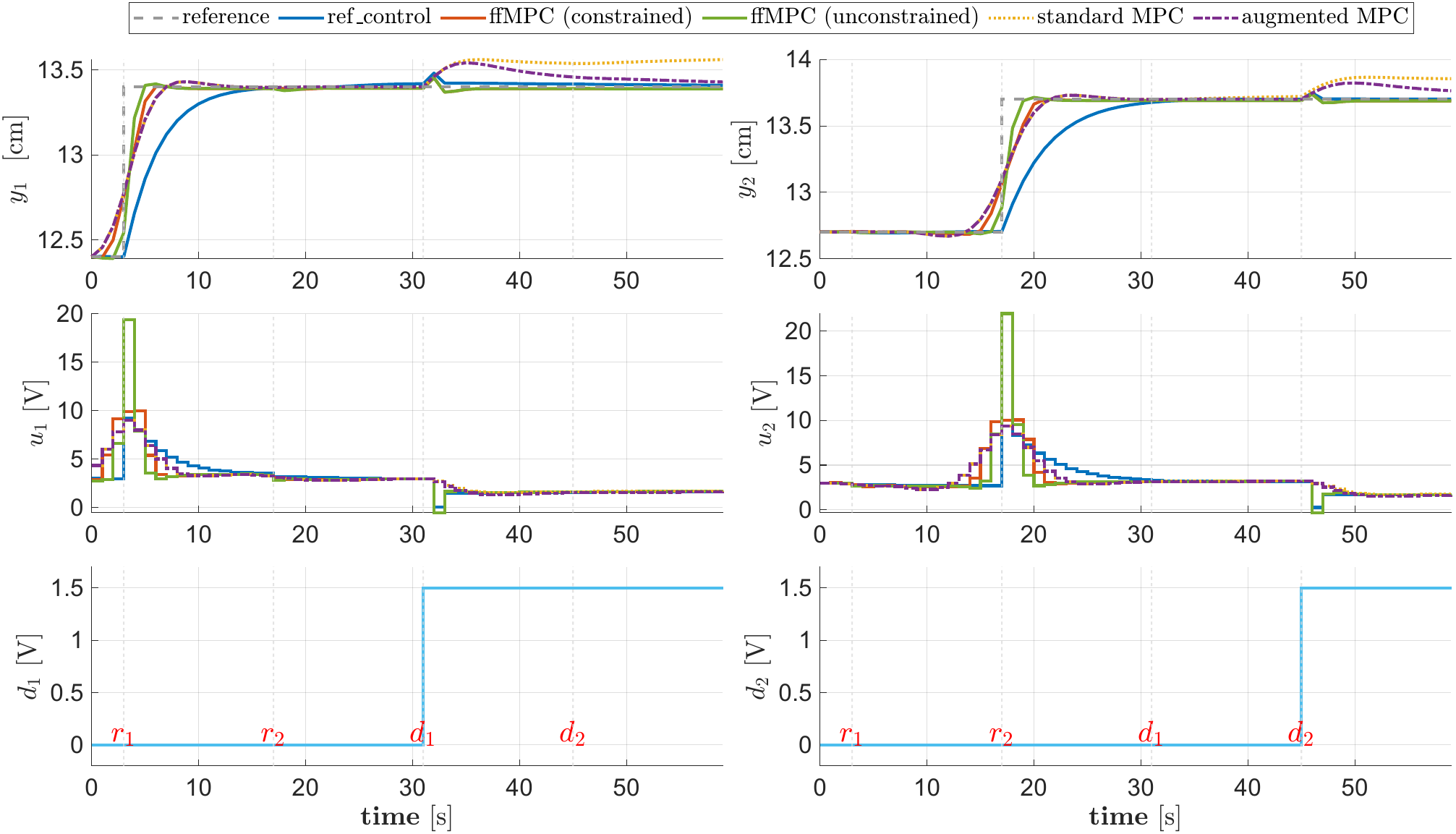}
\caption{Simulation results under different control configurations. ``ref\_control'' represents YK reference control; ``ffMPC (constrained)'' represents YK with feedforward MPC control and constraints; ``ffMPC (unconstrained)'' represents YK with feedforward MPC control without constraints; ``standard MPC'' and ``augmented MPC'' represent traditional MPC control with standard and augmented Kalman filters, respectively.}
\label{fig: all}
\end{figure*}

The simulation result is shown in Figure \ref{fig: all}, where several related cases are considered in simulations. ``ref\_control'' represents the case using the YK control structure with direct reference $r$ imposed without MPC algorithms. ``ffMPC (constrained)'' represents the case using the YK control structure with feedforward MPC considering the constraints meanwhile. ``ffMPC (unconstrained)'' represents the case using YK control structure with feedforward MPC but not implementing constraints. ``standard MPC'' represents the case using traditional MPC control with a non-augmented Kalman filter. ``augmented MPC'' represents the case using traditional MPC control with an augmented Kalman filter, as described in \ref{sec:MPC with augmented Kalman filter}.

The reference tracking performance is evaluated subject to sequential step changes: $r_1$ steps at $t = 3 \text{ [s]}$, followed by $r_2$ at $t = 17 \text{ [s]}$. During these intervals, all frameworks incorporating MPC algorithms exhibit comparable predictive transient dynamics.  Notably, the unconstrained configuration ``ffMPC (unconstrained)'' yields a large transient overshoot in its control signal. Conversely, by explicitly enforcing a actuator limit of $10 \text{ [V]}$, the constrained formulation ``ffMPC (constrained)'' strictly restricts its control effort within the domain ($u \le 10 \text{ [V]}$), successfully mitigating the overshoot without compromising tracking accuracy.

To evaluate the disturbance rejection capabilities, unmeasured step disturbances are injected at $t = 31$ s (for $d_1$) and $t = 45$ s (for $d_2$). Under these, traditional MPC algorithms exhibit markedly inferior performance compared to the proposed hierarchical YK control structure. Specifically, the ``standard MPC'' case inherently fails to compensate for the exogenous inputs, resulting in a prominent steady-state offset. While the ``augmented MPC'' case successfully eliminates this tracking error via its augmented integrator, its transient attenuation is significantly more sluggish than the YK control structure cases, which rapidly neutralize the mismatches.

Since the simulation is conducted on the nonlinear four-tank process, inherent model mismatches arise from applying a linear control strategy to a nonlinear plant. It is worth noting that the discrepancies between the nonlinear system and its linear approximation can be systematically encapsulated by the dual YK parameter $S$. For instance, considering the actual plant operating at a shifted linearization point, the updated dynamic model is represented as:
\begin{equation}
G_{act}:\left[\begin{array}{c|cccc}
A_{act} & B_{act}
\\ \hline
C_{act} & D_{act} 
\end{array}\right]
\end{equation}
where $A_{act}$, $B_{act}$, $C_{act}$, and $D_{act}$ characterize the system dynamics at this new operating point. Its corresponding dual YK parameter can be computed as follows \cite{dualYoulaPara}:
\begin{equation}
    S = V^{-1}V_{act}(\tilde N_{act} M - \tilde M_{act} N)
\end{equation}
where $V_{act}$, $M_{act}$ and $N_{act}$ are the stable coprime factorization results for $G_{act}$ and nominal controller $K_n$. Consequently, as long as $S$ remains stable, the specifically designed parameter $Q$ actively compensates for the mismatch.

\section{Conclusions}
\label{sec:Conclusions}


In this paper, a novel approach for designing a hierarchical 2-degree-of-freedom (2DOF) control architecture combining YK parameterization and MPC is presented. The proposed framework provides a systematic methodology to structurally embed the predictive optimization of MPC within the YK parameterization. By employing an $\mathcal{H}_2$-synthesized YK parameter, the closed-loop system intrinsically achieves offset-free tracking and effectively compensates for inherent model mismatches. Furthermore, this framework fundamentally decouples the MPC component from the stabilizing baseline, enabling it to operate as an independent modular optimizer.




\section*{APPENDIX}
For general state-space models of plant $G_g$ and controller $K_g$ in the form of:
\begin{equation}
    G_g:\left[\begin{array}{c|c}
        A & B
        \\ \hline 
        C & D
\end{array}\right]
\qquad
K_g : 
\left[\begin{array}{c|c}
        \check A & \check B
        \\ \hline 
        \check C & \check D
\end{array}\right]
\end{equation}
The coprime factorization of general state-space models is given by \cite{tay_high_1998m}:
\begin{subequations}\label{eq:coprime fac}
\begin{gather}
\label{eq:coprime1}
    \scalebox{0.7}{$\begin{bmatrix}
    M & U\\
    N & V
    \end{bmatrix}$}
    :
\left[\begin{array}{cc|cc}
A + BF & 0 & B & 0 \\
0 & \check{A} + \check{B}\check{F} & 0 & \check{B} \\ \hline 
F & \check{C} + \check{D}\check{F} & I & \check{D} \\
C + DF & \check{F} & D & I
\end{array}\right]
\\
\label{eq:coprime2}
\scalebox{0.7}{$\begin{bmatrix}
    \tilde{V} & -\tilde{U}\\
    -\tilde{N} & \tilde{M}
    \end{bmatrix}$}
    :
\left[\begin{array}{cc|cc}
A + BY\check{D}C & -BY\check{C} & -BY & BY\check{D} \\
-\check{B}ZC & \check{A} + \check{B}ZD\check{C} & \check{B}ZD & -\check{B}Z \\ \hline 
F - Y\check{D}C & Y\check{C} & Y & -Y\check{D} \\
ZC & \check{F} - ZD\check{C} & -ZD & Z
\end{array}\right]
\\
Y = (I-\check{D}D)^{-1}
        \\
        Z = (I-D\check{D})^{-1}
\end{gather}
\end{subequations}
where $F$ and $\check F$ are the ancillary gain matrices for stabilizing the plant and controller, respectively.


\bibliographystyle{IEEEtran}
\bibliography{manuel_bib}

\end{document}